\documentclass[12pt,reqno]{article}
\usepackage{amsfonts}
\usepackage{amsmath}
\usepackage{amsbsy}

\usepackage{amssymb,latexsym}
\numberwithin{equation}{section}

\begin{document}

\title{Dualisation of the General Scalar Coset in Supergravity Theories}
\author{Nejat T. Y$\i$lmaz\\
          Department of Physics\\
          Middle East Technical University\\
          06531 Ankara,Turkey\\
          \texttt{ntyilmaz@metu.edu.tr}}
\maketitle
\begin{abstract}
The dualised formulation of the symmetric space sigma model is peformed
 for a general scalar coset $G/K$ where $G$ is a maximally non-compact
 group and $K$ it's maximal compact subgroup.By using the twisted self-duality
 condition the general form of the first-order equations
 are obtained.The results are applied to the example of $SL(2,\mathbb{R)}$ /$SO(2)$
 scalar manifold of the IIB supergravity.

\end{abstract}

\section{Introduction}

In supergravity theories which possess scalar fields the global symmetries of the
 scalar sector are essential to have a deeper understanding of these theories.The scalar
 Lagrangians can be formulated as coset sigma models which are non-linear realisations.
The dimension of the coset space is equal to the number of the scalars of the theory
 When we use the Kaluza-Klein dimensional reduction for D=11 supergravity
[1] and obtain the lower dimensional
 maximal supergravity theories the global (rigid) symmetry group
 (i.e.a real Lie group) of the Bosonic sector of the reduced Lagrangians
 are in split real form (maximally non-compact whose Cartan subalgebras
 can be chosen along the non-compact directions).This global symmetry is also the
 global symmetry of the scalars of the theory.The scalar coset can be constructed as
 $G/K$-valued fields where $G$ is the global symmetry group in split real form and
 $K$ it's maximal compact subgroup.The scalars transform non-linearly in the linear representations
 of $G$ while the higher rank potentials always transform linearly.Although the global
 symmetry scheme of the supergravity theories are consequences of
 supersymmetry the study of the coset symmetries of the supergravity theories provides a better
 understanding of the underlying structure of these theories [2,3].

 The method of non-linear realisations [4,5,6,7,8] is used in [9,10]
 to formulate the gravity as a non-linear realisation in which the gravity
 and the gauge fields appear on equal footing.Recently this formulation
 is enlarged to have a detailed treatment of the maximal supergravity theories.

 By introducing auxilary fields for a subset of the field content and
 by using the coset formulation the global symmetries of the scalar sectors
 of maximal supergravities are studied in [11].These symmetries
 which are realised on the Bosonic fields are also studied in a general
 formalism in [12].However the complete coset realisation of the scalars
 and the gauge fields is introduced in [13] where the twisted self-duality
 structure of supergravities [14,15] is generalised to regain the first-order equations
 of the theory from the Cartan forms of the dualised coset.The spacetime symmetries
 turn out to be internal symmetries after the Kaluza-Klein dimensional reduction
 in [11].It is shown that these symmetries become gauge symmetries
 after dualisation.The discussion about the symmetry groups of the Cartan forms (the
 doubled field strengths) of the coset formulation
as well as the symmetry groups of
 the twisted self-duality equation (i.e.first-order equations) which are
 larger than the symmetry groups of the Cartan forms is given in [11] and [13]
.In [13] for IIB theory [16]
 it is also shown that the the global symmetry of the twisted self-duality equation of the
 Cartan form of the coset realisation is the original global symmetry.The same result
 is proved to exist for the dualisation of the scalar sectors of maximal supergravities
 in all dimensions and it is conjectured that the global symmetry is preserved for
 the full dualisation of these theories.

In [17] the complete non-linear realisation of the D=11 and IIA [18,19,20] supergravity
 theories is performed for the entire Bosonic sector including the gravity which is
missing in [13].The symmetries are also discussed in detail in [17].In [21] the
Bosonic sector of IIB is derived as a non-linear realisation.The non-linear
realisations of D=11 supergravity and IIA supergravity lead to finite dimensional Lie
algebras $G_{11}$ and $G_{IIA}$ respectively which are not Kac-Moody algebras.
The local subgroups are also chosen to be the Lorentz group so that the general coset
representative of the entire Bosonic sector can not be parametrised by a Borel
 subgroup of some larger group .These coset formulations are not like the
ones for maximal supergravities which all give rise to Kac-Moody algebras and whose
general coset representatives can be parametrised by a Borel subgroup of a
larger group.In [21,22] it is discussed that the non-linear realisations
of D=11 and IIA supergravities can be enlarged to include a Kac-Moody algebra
identified as $E_{11}$ which contains the Borel subalgebra of $E_{8}$.This can be
done either by introducing a larger local subgroup than the Lorentz group
or by describing the gravity by two fields which are duals of each other.Furhermore
in [21] the IIB theory is formulated as a non-linear realisation of an infinite dimensional
algebra $G_{IIB}$ which is the closure of the conformal algebra.It is also shown that
the non-linear realisation can be enlarged to include the Kac-Moody algebra $E_{11}$
 of IIA and D=11 supergravities.

In this note we will give the dualisation of the general symmetric space coset sigma model
 whose global (rigid) symmetry group is in split real form.The scalar sectors of the maximal
 supergravity theories are in this category and in [11] and [13] the scalar sectors are formulated
case by case with in the context of the dualisation of the Bosonic sectors of these theories.
The formulation presented here assumes a formal construction based on general structure
constants and it is performed for a general $G/K$ scalar manifold ($G$ is a
maximally non-compact group and $K$ is it's maximal compact subgroup).The formulation
is based on the Borel gauge as proposed in [11,13] where the maximal supergravities with
D$\leq 9$ are based on a coset representative in which the axions
appear as various exponential factors.A transformation of the fields is needed
to relate the formulation given here and the one in [11] and [13] for D$\leq 9$.On the other
hand for IIB and IIA we have the exact match of the form of the coset representative
chosen (IIA does not contain axions at all).Among the three possible formulations
of the symmetric space sigma model [11] we will follow the one which
defines an internal metric that is used instead of the coset element.
Although the general form of the coset Lagrangian is given in [13]
the explicit general dualised formulation is not performed.This work completes
the missing general formulation by  presenting the dualisation of the general coset $G/K$ by means of the abstract structure constants.It
also contains the first-order formulation of the equations of motion as a twisted
 self-duality condition for the symmetric space coset sigma model with the global
symmetry group in split real form.

We will start with the derivation of the second-order equations of
motion of the $G/K$ scalar coset manifold in section two.In section three we will calculate
the general structure constants of the dualised algebra in terms of the ones of the
 originally assumed Borel subalgebra by introducing the dual fields and the dual generators.In section four the doubled field strength $\mathcal{G}$ will be
calculated and the first-order equations
will be derived by using the twisted self-duality equation $\ast \mathcal{G=SG}$
 (which is an exact self-duality condition since $\mathcal{S}=1$ in our case).Finally we will give an application of the results we derive for the example
of $SL(2,\mathbb{R)}$ /$SO(2)$ scalar manifold in section five.

\section{The General Scalar Coset Sigma Model}

In this section we will derive the equations of motion for the scalar coset
after we  introduce the Lagrangian.The formulation is parallel with the one
in [23].The equations are second-order and the dualisation will enable us to
obtain the first-order equations of motion.Let $G$ be a maximally
non-compact (split) group and $K$ it's maximal compact subgroup.One can
consider the $G$-valued scalar fields over the D-dimensional spacetime to
form the scalar manifold first and then demand the local $K$ invariance.This
will reduce the degrees of freedom and the physical scalar manifold elements
will be the $G/K$ coset-valued fields.The coset representatives can be
parametrised by the Borel subgroup elements.The Borel subgroup is generated
by the Borel subalgebra whose generators are the Cartan generators \{$%
H_{i}\mid i=1,..,l\}$ of the Cartan subalgebra and the positive root
generators \{$E_{\alpha }\mid \alpha =1,..,m\}$ of the Lie algebra of $G.$So
the $G/K$-valued field $\nu (x)$ can be written as

\begin{equation}
\begin{aligned}
\nu (x)&=\mathbf{g}_{H}(x)\mathbf{g}_{N}(x)\\
\\
&=e^{\frac{1}{2}\phi ^{i}(x)H_{i}}e^{\chi
^{m}(x)E_{m}}.
\end{aligned}
\end{equation}

In this so-called Borel gauge the fields $\{\phi ^{i}\}$ which couple to \{$%
H_{i}\}$ are called the dilatons and the fields \{$\chi ^{m}\}$ which couple
to the positive root generators $\{E_{m}\}$ are called the axions.Totally
there are $n=l+m$ scalar fields which generate the $G/K-$valued fields $\nu
(x)$ of the scalar manifold.

One can parametrise the positive root part of $\nu (x)$ in various ways.For example like in [11,13] the positive root part in (2.1) can be constructed
as products of exponentials of individual positive root generators multiplied with the
corresponding axions.The factors are ordered in a special way so that the axions
chosen are the same fields which appear in the particular form of the Lagrangians used.
The Campbell-Hausdorff formula can be used to relate the field content (dilatons and axions)
of the different parametrisations of $\nu (x)$.Indeed each parametrisation corresponds to a
transformation of the scalar fields but we should also bear in mind that the Borel gauge
of the coset does not change as in all of the parametrisations mentioned above the $G/K$ coset
is still parametrised by the Borel subgroup.

If we define the right,rigid  action of the group $G$ over the $G/K-$valued
fields as $\psi (x)=\nu (x)\mathbf{g}$ for $\mathbf{g}\in G$ we see that $%
\psi (x)$ is not in the Borel gauge .We should introduce another
transformation which would map it into the range of the Borel gauge
again.The Iwasawa decomposion [24] is the tool for this.It assures that any
group element $\mathbf{g}\in $ $G$ which is in the range of the exponential
map\bigskip\ can uniquely be expressed as

\begin{equation}
\mathbf{g}=\mathbf{g}_{K}\mathbf{g}_{H}\mathbf{g}_{N}
\end{equation}

where $\mathbf{g}_{K}\in K$ while $\mathbf{g}_{H}$ and $\mathbf{g}_{N}$ are
the group elements obtained by the exponentiation of the Cartan subalgebra
and the positive root part of the Lie algebra of $G$ respectively.Thus if we
use this result for $\psi (x)$ we have $\psi(x)=\mathbf{g}%
_{K}(x)\nu ^{\prime }(x)$ where $\nu ^{\prime }(x)=\mathbf{g}%
_{K}^{-1}(x)\psi (x)$ is in the Borel gauge.Consequently for each rigid
transformation $\mathbf{g}\in G$ there exists a local transformation $\mathbf{g}%
_{K}^{-1}(x)\in K$ which enables the map

\begin{equation}
  \nu ^{\prime }(x)=\mathbf{g}_{K}^{-1}(x)\nu (x)\mathbf{g}
\end{equation}

to be closed over the range of the Borel gauge.The general scalar coset
Lagrangian is given as [11,13,23]

\begin{equation}
 \mathcal{L}=\frac{1}{4}tr(\ast d\mathcal{M}^{-1}\wedge d\mathcal{M}).
\end{equation}

Here the internal metric $\mathcal{M}$ is $\mathcal{M=}\nu ^{\#}\nu $ where ( \ )$^{\#}$ is the generalised
transpose.In order to define it we first introduce $\tau :g\rightarrow g$
namely the Cartan involution which acts on $g$ the Lie algebra of $G.$It
reverses the sign of the non-compact generators while leaving the compact
generators unchanged.In terms of the basis we use

\begin{equation}
 \tau : (E_{\alpha },E_{-\alpha },H_{i})\longrightarrow (-E_{-\alpha
},-E_{\alpha },-H_{i})
\end{equation}

where $\{E_{-\alpha }\}$ are the negative root generators.For any real group the
compact generators can be written as ($E_{\alpha }-E_{-\alpha })$ and the
non-compact ones as ($E_{\alpha }+E_{-\alpha }).$For a Lie algebra element $%
g^{\prime }$ the general transpose \# is defined as ($g^{\prime
})^{\#}=-\tau (g^{\prime }).$As mentioned in [23] it is possible to find a
higher dimensional representation of the Lie algebra in which \# coincides
with the matrix transpose operator.For this reason one can define a
conjugate \# map over the group $G$ as (exp$g^{\prime })^{\#}=$ exp($%
g^{\prime \#}).$If the subgroup of $G$ generated by the compact
generators is an orthogonal group then in the fundamental
representation the generators
can be chosen such that $\mathbf{g}^{\#}=\mathbf{g}^{\intercal } ($i.e.$SL(n,\mathbb{R))}$.If it is a unitary group then $\mathbf{g}^{\#}=\mathbf{g}%
^{\dagger }.$Therefore the generalized transpose of the coset
representative is defined as follows

\begin{equation}
\begin{aligned}
\nu ^{\#}&=\tau (\nu ^{-1})\\
\\
&= e^{\chi ^{m}E_{-m}}e^{\frac{1}{2}\phi
^{i}H_{i}}.
\end{aligned}
\end{equation}

The Iwasawa decomposition assures that the Lagrangian (2.4) is invariant
under the rigid transformations of $G$ which we have defined before.As
discussed in [13] the transformation acting on the general Noether currents
generated by the rigid (non-coordinate dependent) right action of $G$ on the
scalar coset manifold is linear.Since the number of currents exceeds the
total number of scalars (dilatons and axions) of the theory these currents
are not all independent.Therefore the Borel currents corresponding to the
symmetries generated by the Borel subalgebra transform non-linearly under
the action of $G.$This may be phrased as the non-linear realisation of the
global symmetry group $G$ over the scalars.

By using $\nu ^{-1}d\nu=-d\nu ^{-1}\nu $ and the
properties of the generalized transpose \# also the fact that the
cyclic permutations are permissable under the trace the scalar
coset Lagrangian can be expressed as

\begin{equation}
 \mathcal{L=-}\frac{1}{2}tr(\ast d\nu \nu ^{-1}\wedge (d\nu \nu
^{-1})^{\#}+\ast d\nu \nu ^{-1}\wedge d\nu \nu ^{-1}).
\end{equation}

As mentioned in section one the scalar coset manifolds of the maximal supergravity
theories can be given as $G/K$ ,$G$ being a maximally non-compact group and $K$ it's
maximal compact subgroup.For D=11 supergravity the global symmetry group
$G$ is the trivial group,for IIB $G$ is
 $SL(2,\mathbb{R)}$ and $K$ is $SO(2)$,for IIA $G$ is $SO(1,1)/\mathbb{Z}
_{2}$ and $K$ is the trivial group.The D=9 maximal supergravity has the scalar coset
 $GL(2,\mathbb{R)}$/$SO(2)$.Exceptional groups $E_{n}$ arise  as the global symmetry groups
for $D<9$ maximal
supergravities.The details of the maximal compact subgroups of these theories
 can be found in [24].

The 1-form $\mathcal{G}_{0}=d\nu \nu ^{-1}$ is the pullback of a Lie
algebra valued 1-form on $G$ through the map $\nu .$For this reason it can
be expressed in the Borel subalgebra basis

\begin{equation}
\begin{aligned}
\mathcal{G}_{0}&=d\nu \nu ^{-1}\\
\\
&=(d\mathbf{g}_{H}\mathbf{g}_{N}+\mathbf{g}_{H}d\mathbf{g}%
_{N})(\mathbf{g}_{N}^{-1}\mathbf{g}_{H}^{-1})\\
\\
&=d\mathbf{g}_{H}\mathbf{g}_{H}^{-1}+\mathbf{g}_{H}d\mathbf{g}_{N}\mathbf{g}_{N}^{-1}\mathbf{g}_{H}^{-1}.
\end{aligned}
\end{equation}

Now in order to simplify the first term we will use the formula

\begin{equation}
de^{X}e^{-X}=dX+\frac{1}{2!}[X,dX]+\frac{1}{3!}[X,[X,
dX]]+....
\end{equation}

Therefore we have

\begin{equation}
\begin{aligned}
d\mathbf{g}_{H}\mathbf{g}_{H}^{-1}&= de^{\frac{1}{2}\phi
^{i}H_{i}}e^{-\frac{1}{2}\phi ^{i}H_{i}}\\
\\
&=\frac{1}{2}d\phi ^{i}H_{i}
\end{aligned}
\end{equation}

where we have used the fact that $[H_{i},H_{j}]=0.$The other commutation
relations of the Borel subalgebra are $[H_{i},E_{\alpha }]=\alpha
_{i}E_{\alpha }$ and $[E_{\alpha },E_{\beta }]=N_{\alpha ,\beta }E_{\alpha
+\beta }.$By using the above expansion for $de^{X}e^{-X}$ we can also calculate $d\mathbf{g}_{N}\mathbf{g}_{N}^{-1}$

\begin{equation}
\begin{aligned}
d\mathbf{g}_{N}\mathbf{g}_{N}^{-1}&=de^{\chi ^{m}E_{m}}e^{-\chi ^{m}E_{m}}\\
\\
&=d\chi ^{m}E_{m}+\frac{1}{2!}[\chi ^{m}E_{m},d%
\chi ^{n}E_{n}]+\frac{1}{3!}[\chi ^{m}E_{m},[\chi ^{l}E_{l},d\chi
^{n}E_{n}]]+....\\
\\
&=d\chi ^{m}E_{m}+\frac{1}{2!}\chi ^{m}d\chi
^{n}K_{mn}^{v}E_{v}+\frac{1}{3!}\chi ^{m}\chi ^{l}d\chi
^{n}K_{ln}^{v}K_{mv}^{u}E_{u}+....\\
\\
&=\overset{\rightharpoonup}{\mathbf{E
}}\:\mathbf{\Omega }\:\overset{\rightharpoonup }{d\chi }.
\end{aligned}
\end{equation}

In the compact form on the right hand side of the last line $\overset{
\rightharpoonup }{\mathbf{E}}$ is the row vector of the
positive root generators; $\mathbf{E}_{\alpha }=E_{\alpha }$ and $\overset{%
\rightharpoonup }{d\chi }$ is the column vector of the 1-forms $\{d\chi
^{\alpha }\}.$The matrix $\mathbf{\Omega }$ is a series which arises from
the infinite sum given above

\begin{equation}
\mathbf{\Omega}=\sum\limits_{n=0}^{\infty }\dfrac{\omega ^{n}}{(n+1)!}
\end{equation}

$\omega $ is an $m\times m$ matrix $m$ being the number of the positive
roots and it is composed of the axions coupled to the structure constants $%
\omega _{\beta }^{\gamma }=\chi ^{\alpha }\,K_{\alpha \beta }^{\gamma }.$In
this definition we have introduced [$E_{\alpha },E_{\beta }]=K_{\alpha \beta
}^{\gamma }\,E_{\gamma }.$In other words $K_{\beta \beta }^{\alpha
}=0,K_{\beta \gamma }^{\alpha }=N_{\beta ,\gamma }$ if in the root sense $%
\beta +\gamma =\alpha $ and $K_{\beta \gamma }^{\alpha }=0$ if $\beta
+\gamma \neq \alpha $ again in the root sense.Now if we use the
Campbell-Hausdorff formula

\begin{equation}
e^{X}Ye^{-X}=Y+[X,Y]+\frac{1}{2!}[X,[X,Y]]+....
\end{equation}

we can explicitly write the second term in (2.8) as

\begin{equation}
\begin{aligned}
\mathbf{g}_{H}d\mathbf{g}_{N}\mathbf{g}_{N}^{-1}\mathbf{g}_{H}^{-1}&=%
e^{\frac{1}{2}\phi ^{i}H_{i}}(\overset{\rightharpoonup }{\mathbf{E}}%
\:\mathbf{\Omega }\:\overset{\rightharpoonup }{d\chi })e^{-\frac{1%
}{2}\phi ^{i}H_{i}}\\
\\
&=\overset{\rightharpoonup }{
\mathbf{E}^{\prime }}\:\mathbf{\Omega }\:\overset{\rightharpoonup }{d\chi
}.
\end{aligned}
\end{equation}

The primed row vector $\overset{\rightharpoonup }{\mathbf{E}^{\prime }}$ is
defined as $\mathbf{E}^{\prime}_{\beta }=e^{\frac{1}{2}\beta _{i}\phi
^{i}}E_{\beta }.$Finally we can now write $\mathcal{G}_{0}$ expanded in the
Borel subalgebra generators

\begin{equation}
\begin{aligned}
\mathcal{G}_{0}&=d\nu \nu ^{-1}\\
\\
&=\frac{1}{2}d\phi ^{i}H_{i}+e^{%
\frac{1}{2}\alpha _{i}\phi ^{i}}F^{\alpha }E_{\alpha }
\end{aligned}
\end{equation}

We have also introduced the vector $F^{\alpha }=\mathbf{\Omega }_{\beta
}^{\alpha }\,d\chi ^{\beta }.$We will express the equations of motion in
terms of $\overset{\rightharpoonup }{\mathbf{F}}.$If we use \
(2.15) in (2.7) we find that

\begin{equation}
\mathcal{L=}-\frac{1}{2}\sum\limits_{i=1}^{l}\ast d\phi ^{i}\wedge d%
\phi ^{i}-\frac{1}{2}\sum\limits_{\alpha =1}^{m}e^{\alpha _{i}\phi
^{i}}\ast F^{\alpha }\wedge F^{\alpha }.
\end{equation}

By following the outline of [23] one can derive the equations of motion of
the general scalar coset Lagrangian.We should first observe that $ d(d\nu \nu
^{-1})=-d\nu \wedge d\nu ^{-1}=d\nu \nu ^{-1}\wedge d\nu
\nu ^{-1}.$If (2.15) is substituted into this equation one gets the Bianchi
identity for $\overset{\rightharpoonup }{\mathbf{F}}.$

\begin{equation}
dF^{\gamma }=\frac{1}{2}\sum\limits_{\alpha +\beta =\gamma }N_{\alpha
,\beta }F^{\alpha }\wedge F^{\beta }.
\end{equation}

If $\{F^{\gamma }\}$ are considered as independent fields and if we propose the
Bianchi identity as a constraint equation the Lagrange multipliers ( $(D
-2) $-forms) can be introduced and the additional Lagrangian corresponding
to the Bianchi identity can be given as

\begin{equation}
\mathcal{L}_{Bianchi}=(dF^{\alpha }-\frac{1}{2}\sum\limits_{\alpha =\beta
+\gamma }N_{\beta ,\gamma }F^{\beta }\wedge F^{\gamma })\wedge A_{(
D-2),\alpha }.
\end{equation}

The new Lagrangian becomes $\mathcal{L}^{\prime }=\mathcal{L}+\mathcal{L}_{Bianchi}.$%
The variation with respect to $A_{(D-2),\alpha }$ for $\alpha
=1,..,m$ will give back the Bianchi identities.If we vary $\mathcal{L}%
^{\prime }$ with respect to $F^{\gamma }$ and then take the exterior
derivative of the resulting field equation we achieve the second-order
equations of motion for $F^{\gamma }$

\begin{equation}
d(e^{\gamma _{i}\phi ^{i}}\ast F^{\gamma })=\sum\limits_{\alpha
-\beta =-\gamma }N_{\alpha ,-\beta }F^{\alpha }\wedge e^{\beta
_{i}\phi ^{i}}\ast F^{\beta }.
\end{equation}

By varying (2.16) with respect to the dilatons $\{\phi ^{i}\}$ (since $\mathcal{L}_{Bianchi}
$ does not depend on $\{\phi ^{i}\}$) we can also find the equations of motion
for $\phi ^{i}$

\begin{equation}
d(\ast d\phi ^{i})=\frac{1}{2}\sum\limits_{\alpha =1}^{m}\alpha _{i}%
e^{\frac{1}{2}\alpha _{i}\phi ^{i}}F^{\alpha }\wedge e^{%
\frac{1}{2}\alpha _{i}\phi ^{i}}\ast F^{\alpha }.
\end{equation}

The details of the formulation given above can be found in [23].As we will
see in the next section in order to formulate the dualised theory we need to
use a slightly different form of (2.19) namely

\begin{equation}
\begin{aligned}
d(e^{\frac{1}{2}\gamma _{i}\phi ^{i}}\ast F^{\gamma })&=d(e
^{-\frac{1}{2}\gamma _{i}\phi ^{i}}e^{\gamma _{i}\phi ^{i}}\ast
F^{\gamma })\\
\\
&=-\frac{1}{2}\gamma _{j}e^{-\frac{1}{2}\gamma _{i}\phi ^{i}}d\phi ^{j}\wedge e
^{\gamma _{i}\phi ^{i}}\ast F^{\gamma }\\
\\
&+\sum\limits_{\alpha -\beta =-\gamma }
e^{-\frac{1}{2}\gamma _{i}\phi ^{i}}N_{\alpha ,-\beta }F^{\alpha
}\wedge e^{\beta _{i}\phi ^{i}}\ast F^{\beta }\\
\\
&=-\frac{1}{2}\gamma _{j}e^{\frac{1}{2}\gamma _{i}\phi ^{i}}d\phi ^{j}\wedge \ast
F^{\gamma }+\sum\limits_{\alpha -\beta =-\gamma }e^{\frac{1}{2}
\alpha _{i}\phi ^{i}}e^{\frac{1}{2}\beta _{i}\phi ^{i}}N_{\alpha
,-\beta }F^{\alpha }\wedge \ast F^{\beta }.
\end{aligned}
\end{equation}

The second-order equations (2.20) and (2.21) are the ones which we will
refer to derive the commutation relations of the dualised generators when we
construct the doubled field strength $\mathcal{G}$ which will give the
correct first-order equations by satisfying a twisted self-duality condition.

\section{Dualisation}

If we double the number of the fields by introducing a $(D-2)$-form for
each scalar field we can construct a doubled field strength $\mathcal{G}$
which is Lie superalgebra valued [11,13].This algebra is generated by the
Borel generators and the generators we introduce for each dual $(D-2$
)-form whose commutation relations we will calculate.As we will see
the choice of the appropriate structure constants will be based on the formulation which will give the correct equations of motion.The doubled field strength $\mathcal{G}$ is invariant under the Borel
subgroup $G_{+}$ of $G$ which corresponds to the rigid symmetries of the
constant shifts of the dilatons and the axions and $\mathcal{G}$ is also invariant under the
local gauge symmetries which are generated by the generators coupled to the
dual $(D-2)$-forms as discussed in [13].For the doubled field content we
propose the map $\nu ^{\prime }(x)$ as

\begin{equation}
\nu ^{\prime }(x)=e^{\frac{1}{2}\phi ^{i}H_{i}}e^{\chi
^{m}E_{m}}e^{\widetilde{\chi }^{m}\widetilde{E}_{m}}e^{
\frac{1}{2}\widetilde{\phi }^{i}\widetilde{H}_{i}}.
\end{equation}

The doubled field strength $\mathcal{G}$ is defined as $d\nu ^{\prime }\nu
^{\prime -1}.$In [11] and [13] the twisted self-duality condition is applied
on $\mathcal{G}$ to regain the first-order equations of the maximal
supergravity theories.The twisted self-duality condition which is imposed on
$\mathcal{G}$ is $\ast \mathcal{G=SG}.$Here $\mathcal{S}$ is a
pseudo-involution of the proposed Lie superalgebra (which is generated by
the original Borel generators and their duals).It's action on the scalar and
dual generators is as follows

\begin{gather}
\mathcal{S}H_{i}=\widetilde{H}_{i}\quad,\quad\mathcal{S}E_{m}=%
\widetilde{E}_{m}, \notag\\
\notag\\
\mathcal{S}\widetilde{H}_{i}=H_{i}\quad,\quad\mathcal{S}%
\widetilde{E}_{m}=E_{m}.
\end{gather}

In general $\mathcal{S}$ sends the Borel generators to their duals and the
dual ones to their Borel counterparts with a sign factor which is $(-1)^{p(%
D-p)+s}$where $p$ is the degree of the corresponding field strength
which the generator is coupled to and $s$ is the signature of the spacetime
metric.In fact the eigenvalues of $\mathcal{S}^{2}$ must be the same with
the eigenvalues of $(\ast \circ \ast )$ operator acting on the field
strength coupled to the dual generator which $\mathcal{S}^{2}$ acts on.For
the present case the field strengths are $(D-1)$-forms so $p=D-1$ and
the signature of the spacetime is assumed to be $s=D-1.$So the sign factor
is  $(-1)^{p(D-p)+s}=(-1)^{2(D-1)}=1.$

The Borel generators satisfy the commutation relations given in the previous
section and as mentioned in [13] the general form of the remaining
commutation relations of the newly constructed algebra are

\begin{gather}
[E_{\alpha },\widetilde{T}_{m}]=\widetilde{f}_{\alpha m}^{n}\widetilde{T}
_{n}\quad, \quad [H_{i},\widetilde{T}_{m}]=\widetilde{g}_{im}^{n}
\widetilde{T}_{n},\notag\\
\notag\\
[\widetilde{T}_{m},\widetilde{T}_{n}]=0
\end{gather}

where $\widetilde{T}_{i}=\widetilde{H}_{i}$ for $i=1,...,l$ and $\widetilde{T%
}_{\alpha +l}=\widetilde{E}_{\alpha }$ for $\alpha =1,...,m.$In the next
section we will calculate the doubled field strength $\mathcal{G=}d\nu
^{\prime }\nu ^{\prime -1}$ explicitly but the following formulation will
make use of an alternative method introduced and used in [13].We will define
a the field strength $\mathcal{G}^{\prime }$ as $\mathcal{G}^{\prime }=%
\mathcal{G}_{0}+\mathcal{S}\ast \mathcal{G}_{0}.$By using (2.15)

\begin{equation}
\mathcal{G}^{\prime }=d\nu \nu ^{-1}+\frac{1}{2}\ast d\phi ^{i}%
\widetilde{H}_{i}+e^{\frac{1}{2}\alpha _{i}\phi ^{i}}\ast F^{\alpha
}\widetilde{E}_{\alpha }.
\end{equation}

From it's definition $\mathcal{G}^{\prime }$ trivially satisfies the twisted
self-duality condition $\ast \mathcal{G}^{\prime }\mathcal{=SG}^{\prime }.$%
Since $d(d\nu \nu ^{-1})=$ $d\nu \nu ^{-1}\wedge d\nu \nu ^{-1}$ $%
\mathcal{G}_{0}$ satisfies the Cartan-Maurer equation $d\mathcal{G}_{0}-%
\mathcal{G}_{0}\wedge \mathcal{G}_{0}=0.$As a matter of fact $\mathcal{G}^{\prime }$ is
nothing but $\mathcal{G}$ in which the self-duality condition $\ast \mathcal{%
G=SG}$ is used for this reason $\mathcal{G}^{\prime }$ also satisfies the
Cartan-Maurer equation

\begin{equation}
d\mathcal{G}^{\prime }-\mathcal{G}^{\prime }\wedge \mathcal{G}^{\prime }=0.
\end{equation}

This equation gives the original second-order equations of the coset
Lagrangian.Therefore one way of obtaining the structure constants in (3.3)
is to calculate (3.5) by substituting (3.4) and to compare the result with
the second-order equations (2.20) and (2.21).

We will use this fact to read the structure constants in (3.3) in terms of
the information of the Borel subalgebra of $G.$We should bear in mind that
the generators $H_{i}$ and $E_{\alpha }$ are of even degree and the ones $%
\widetilde{H}_{i}$ and $\widetilde{E}_{\alpha }$ are of even or odd degree
whether their corresponding potentials $(D-2)$-forms are even or odd rank
depending on the spacetime dimension.In the exterior algebra of the Lie
superalgebra valued forms even or odd generators behave like even or odd
degree forms when commuting with the differential forms.Also if $T$ is an
odd generator $d(TA)=-TdA$ and if it is even $d(TA)=TdA.$Now by using
(3.4) if we calculate d$\mathcal{G}^{\prime }-\mathcal{G}^{\prime }\wedge
\mathcal{G}^{\prime }$ and equate it to zero we find that

\begin{equation}
\begin{aligned}
d\mathcal{G}^{\prime }-\mathcal{G}^{\prime }\wedge \mathcal{G}^{\prime }&=%
\frac{1}{2}d(\ast d\phi ^{i})\widetilde{H}_{i}+d(e^{\frac{1}{2%
}\alpha _{i}\phi ^{i}}\ast F^{\alpha })\widetilde{E}_{\alpha }-\frac{1}{4}d%
\phi ^{j}\wedge \ast d\phi ^{i}[H_{j},\widetilde{H}_{i}]\\
\\
&-\frac{1}{2}d\phi ^{j}\wedge e^{\frac{1}{2}\alpha _{i}\phi
^{i}}\ast F^{\alpha }[H_{j},\widetilde{E}_{\alpha }]-\frac{1}{2}e^{%
\frac{1}{2}\alpha _{i}\phi ^{i}}F^{\alpha }\wedge \ast d\phi
^{j}[E_{\alpha },\widetilde{H}_{j}] \\
\\
&-e^{\frac{1}{2}\alpha _{i}\phi ^{i}}e^{\frac{1}{2}\beta
_{i}\phi ^{i}}F^{\alpha }\wedge \ast F^{\beta }[E_{\alpha },\widetilde{E}%
_{\beta }].
\end{aligned}
\end{equation}

We have to equate the coefficients of the linearly independent algebra
generators to zero and the resulting equations must be the same with the
second-order equations of motion (2.20) and (2.21).Direct comparison gives
the commutation relations as

\begin{gather}
[H_{j},\widetilde{H}_{i}]=0\quad\quad,\quad\quad [E_{\alpha },\widetilde{%
H}_{j}]=0,\notag\\
\notag\\
[H_{j},\widetilde{E}_{\alpha }]=-\alpha _{j}\widetilde{E}_{\alpha }\quad ,\quad [%
E_{\alpha },\widetilde{E}_{\alpha }]=\frac{1}{4}\overset{l}{\underset{i=1}{%
\sum }}\alpha _{i}\widetilde{H}_{i},\notag\\
\notag\\
[E_{\alpha },\widetilde{E}_{\beta }]=N_{\alpha ,-\beta }\widetilde{E}%
_{\gamma },\quad\quad\alpha -\beta =-\gamma,\alpha \neq \beta.
\end{gather}

The conditions in the last line must be understood in the root sense.We can
now express the structure constants $\widetilde{f}_{\alpha m}^{n}$ and $%
\widetilde{g}_{im}^{n}$ by using the results we have found

\begin{gather}
\widetilde{f}_{\alpha m}^{n}=0,\quad\quad m\leq l\quad,\quad
\widetilde{f}_{\alpha ,\alpha +l}^{i}=\frac{1}{4}\alpha _{i},\quad\quad%
i\leq l\notag\\
\notag\\
\widetilde{f}_{\alpha ,\alpha +l}^{i}=0,\quad\quad i>l\quad,\quad
\widetilde{f}_{\alpha ,\beta +l}^{i}=0,\quad\quad i\leq l,%
\alpha \neq \beta \notag\\
\notag\\
\widetilde{f}_{\alpha ,\beta +l}^{\gamma +l}=N_{\alpha ,-\beta },\quad\quad \alpha -\beta =-\gamma\notag\\
\notag\\
\widetilde{f}_{\alpha ,\beta +l}^{\gamma +l}=0,\quad\quad \alpha -\beta
\neq -\gamma ,\alpha \neq \beta .
\end{gather}

The conditions on the indices $\alpha$ and $\beta$ are in the root sense.Also we have

\begin{gather}
\widetilde{g}_{im}^{n}=0,\quad\quad m\leq l\quad,\quad\widetilde{%
g}_{im}^{n}=0,\quad\quad m>l,m\neq n\notag\\
\notag\\
\widetilde{g}_{i\alpha }^{\alpha }=-\alpha _{i},\quad\quad\alpha >l.
\end{gather}

\section{The First-Order Equations of Motion}

Now that we have obtained the complete commutation relations of the dualised
algebra we can explicitly calculate the doubled field strength in terms of
the structure constants and then use the twisted self-duality equation $\ast
\mathcal{G=SG}$ to reach the first-order equations of motion.We should make
the remark that although the symmetry group of $\mathcal{G}$ is composed of
the Borel subgroup $G_{+}$ and the local gauge symmetries of the dual
generators the symmetry group which leaves the first-order equations $\ast
\mathcal{G=SG}$ is larger than that.The reason for this is that there could
be symmetry transformations which may change $\mathcal{G}$ but still leave
the twisted self-duality condition invariant.By using (3.1) we have

\begin{equation}
\begin{aligned}
\mathcal{G=}d\nu ^{\prime }\nu ^{\prime -1}&=\mathcal{G}_{0}+e^{%
\frac{1}{2}\phi ^{i}H_{i}}e^{\chi ^{m}E_{m}}de^{%
\widetilde{\chi }^{m}\widetilde{E}_{m}}e^{-\widetilde{\chi }^{m}%
\widetilde{E}_{m}}e^{-\chi ^{m}E_{m}}e^{-\frac{1}{2}\phi
^{i}H_{i}}\\
\\
&+e^{\frac{1}{2}\phi ^{i}H_{i}}%
e^{\chi ^{m}E_{m}}e^{\widetilde{\chi }^{m}\widetilde{E}%
_{m}}de^{\frac{1}{2}\widetilde{\phi }^{i}\widetilde{H}_{i}}%
e^{-\frac{1}{2}\widetilde{\phi }^{i}\widetilde{H}_{i}}e^{-%
\widetilde{\chi }^{m}\widetilde{E}_{m}}e^{-\chi ^{m}E_{m}}e%
^{-\frac{1}{2}\phi ^{i}H_{i}}.
\end{aligned}
\end{equation}

 Now by using the equation (2.9) and the fact that [$\widetilde{E}_{m},\widetilde{E}_{n}]=[%
\widetilde{H}_{i},\widetilde{H_{j}}]=[\widetilde{E}_{m},\widetilde{H}_{i}]=0$
we obtain

\begin{equation}
\begin{aligned}
de^{\widetilde{\chi }^{m}\widetilde{E}_{m}}e^{-\widetilde{%
\chi }^{m}\widetilde{E}_{m}}&=d\widetilde{\chi }^{m}\widetilde{E}_{m}\\
\\
e^{\widetilde{\chi }^{m}\widetilde{E}_{m}}de^{\frac{1}{2%
}\widetilde{\phi }^{i}\widetilde{H}_{i}}e^{-\frac{1}{2}\widetilde{%
\phi }^{i}\widetilde{H}_{i}}e^{-\widetilde{\chi }^{m}\widetilde{E}%
_{m}}&=\frac{1}{2}d\widetilde{\phi }^{i}\widetilde{H}_{i}.
\end{aligned}
\end{equation}

 So the field strength becomes

\begin{equation}
\mathcal{G=G}_{0}+e^{\frac{1}{2}\phi ^{i}H_{i}}e^{\chi
^{m}E_{m}}A^{n}\widetilde{T}_{n}e^{-\chi ^{m}E_{m}}e^{-%
\frac{1}{2}\phi ^{i}H_{i}}
\end{equation}

where we have defined the vector $\overset{\rightharpoonup }{\mathbf{A}}$ as
$A^{i}=\frac{1}{2}d\widetilde{\phi }^{i}$ for $i=1,...,l$ and $A^{i+\alpha
}=d\widetilde{\chi }^{\alpha }$ for $\alpha =1,...,m.$If we use the
Campbell-Hausdorff formula (2.13) twice we can calculate the second term in
(4.3)

\begin{equation}
\begin{aligned}
e^{\chi ^{m}E_{m}}A^{n}\widetilde{T}_{n}e^{-\chi
^{m}E_{m}}&=A^{n}\widetilde{T}_{n}+[\chi ^{m}E_{m},A^{n}\widetilde{T}_{n}]+%
\frac{1}{2!}[\chi ^{m}E_{m},[\chi ^{l}E_{l},A^{n}\widetilde{T}_{n}]]+....\\
\\
&=A^{n}%
\widetilde{T}_{n}+\chi ^{m}A^{n}\widetilde{f}_{mn}^{k}\widetilde{T}_{k}+%
\frac{1}{2!}\chi ^{m}\chi ^{l}A^{n}\widetilde{f}_{mk}^{v}\widetilde{f}%
_{ln}^{k}\widetilde{T}_{v}+....\\
\\
&=\overset{%
\rightharpoonup }{\widetilde{\mathbf{T}}}\:e^{\mathbf{\Lambda }}%
\:\overset{\rightharpoonup }{\mathbf{A}}.
\end{aligned}
\end{equation}

We have defined the row vector $\overset{\rightharpoonup }{\widetilde{%
\mathbf{T}}}$ in section three and the matrix $\mathbf{\Lambda }$ is defined
as $\mathbf{\Lambda }_{n}^{k}=\chi ^{m}\widetilde{f}_{mn}^{k}.$Now applying
the Campbell-Hausdorff formula (2.13) once more

\begin{equation}
\begin{aligned}
e^{\frac{1}{2}\phi ^{i}H_{i}}(\overset{\rightharpoonup }{%
\widetilde{\mathbf{T}}}\:e^{\mathbf{\Lambda }}\:\overset{%
\rightharpoonup }{\mathbf{A}})e^{-\frac{1}{2}\phi ^{i}H_{i}}&=%
\overset{\rightharpoonup }{\widetilde{\mathbf{T}}}\:e^{\mathbf{%
\Lambda }}\:\overset{\rightharpoonup }{\mathbf{A}}+[\frac{1}{2}\phi ^{i}H_{i},%
(\overset{\rightharpoonup }{\widetilde{\mathbf{T}}}\:e^{\mathbf{%
\Lambda }}\:\overset{\rightharpoonup }{\mathbf{A}})]\\
\\
&+%
\frac{1}{2!}[\frac{1}{2}\phi ^{i}H_{i},[\frac{1}{2}\phi ^{j}H_{j},(\overset{%
\rightharpoonup }{\widetilde{\mathbf{T}}}\:e^{\mathbf{\Lambda }}%
\:\overset{\rightharpoonup }{\mathbf{A}})]]+....\\
\\
&=%
\overset{\rightharpoonup }{\widetilde{\mathbf{T}}}\:e^{\mathbf{%
\Lambda }}\:\overset{\rightharpoonup }{\mathbf{A}}+\frac{1}{2}\phi ^{i}(%
e^{\mathbf{\Lambda }})_{l}^{k}A^{l}\widetilde{g}_{ik}^{v}\widetilde{%
T}_{v}\\
\\
&+\frac{1}{2!}\frac{1}{4}\phi ^{i}\phi ^{j}(e^{\mathbf{\Lambda
}})_{l}^{k}A^{l}\widetilde{g}_{is}^{u}\widetilde{g}_{jk}^{s}\widetilde{T}%
_{u}+....\\
\\
&=%
\overset{\rightharpoonup }{\widetilde{\mathbf{T}}}\:e^{\mathbf{%
\Gamma }}\:e^{\mathbf{\Lambda }}\:\overset{\rightharpoonup }{\mathbf{A}}
\end{aligned}
\end{equation}

where we have defined the matrix $\mathbf{\Gamma }$ as $\mathbf{\Gamma }%
_{n}^{k}=\frac{1}{2}\phi ^{i}\,\widetilde{g}_{in}^{k}$.We can now write the
doubled field strength $\mathcal{G}$

\begin{equation}
\begin{aligned}
\mathcal{G}&=\mathcal{G}_{0}+\overset{\rightharpoonup }{\widetilde{\mathbf{T}}}\:e
^{\mathbf{\Gamma }}\:e^{\mathbf{\Lambda }}\:\overset{\rightharpoonup
}{\mathbf{A}}\\
\\
&=\frac{1}{2}d\phi ^{i}H_{i}+\overset{\rightharpoonup }{\mathbf{E}%
^{\prime }}\:\mathbf{\Omega }\:\overset{\rightharpoonup }{d\chi }+%
\overset{\rightharpoonup }{\widetilde{\mathbf{T}}}\:e^{\mathbf{\Gamma
}}\:e^{\mathbf{\Lambda }}\:\overset{\rightharpoonup }{\mathbf{A}}.
\end{aligned}
\end{equation}

As we have expressed $\mathcal{G}$ in the basis $\{H_{i},E_{\alpha },%
\widetilde{H}_{i},\widetilde{E}_{\alpha }\}$ finally we can find the first
order equations of the scalar Lagrangian (2.7) by using the twisted self-
duality equation $\ast \mathcal{G=SG}$ .If we substitute (4.6) into this
equation we can read the first-order equations for the dilatons and the
axions in terms of the dual $(D-1)$-form field strengths.So the first-order
equations are obtained by equating the coefficients of the  same generators on opposite sides of the equation $%
\ast \mathcal{G=SG}$ and in compact form they can be written as

\begin{equation}
\ast \overset{\rightharpoonup }{\mathbf{\Psi }}=e^{\mathbf{\Gamma }%
}e^{\mathbf{\Lambda }}\overset{\rightharpoonup }{\mathbf{A}}.
\end{equation}

In this vector equation the column vector $\overset{\rightharpoonup }{%
\mathbf{\Psi }}$ is defined as $\Psi ^{i}=\frac{1}{2}d\phi ^{i}$ for $%
i=1,...,l$ and $\Psi ^{\alpha +l}=e^{\frac{1}{2}\alpha _{i}\phi
^{i}}\mathbf{\Omega }_{l}^{\alpha }\,d\chi ^{l}$ for $\alpha =1,...,m$ where
$\mathbf{\Omega }$ is the $m\times m$ matrix defined in the second
section.This concludes our formulation next we will give an application of
the results we have obtained for the $SL(2,\mathbb{R)}$ coset.

\section{The $SL(2,\mathbb{R)}$ /$SO(2)$ Scalar Coset}

The $SL(2,\mathbb{R)}$ /$SO(2)$ scalar coset Lagrangians arise when the
Kaluza-Klein reduction is applied over Tori.For example when the pure
gravity in D+2 dimensions is reduced to D dimensions over the 2-Torus T$^{2}$
the \ global symmetry of the reduced scalar Lagrangian becomes $\mathbb{%
R\times }SL(2,\mathbb{R)}$.The Scalar coset manifold of IIB supergravity
also appears as $SL(2,\mathbb{R})$ /$SO(2)$.The dualisation of
the  $SL(2,\mathbb{R)}$ /$SO(2)$ coset
is studied in [11,13].We will derive the same results by using
the general formulation we have developed.The Lie algebra of $SL(2,\mathbb{R)%
}$ namely $sl(2,\mathbb{R)}$ is isomorphic to $su(2)$ and it has three generators \{$H,E_{+},E_{-}\}$ so
it's Cartan subalgebras are one dimensional and there is one positive root
and one negative root.Therefore the Borel subalgebra is generated by \{$%
H,E_{+}\}$ and they satisfy the commutation relation [$H,E_{+}]=2E_{+}$
denoting $\alpha =2.$For $sl(2,\mathbb{R)}$ one can chose the following
representation

\begin{equation}
H=\left(
\begin{array}{cc}
1 & 0 \\
0 & -1
\end{array}
\right) \quad E_{+}=\left(
\begin{array}{cc}
0 & 1 \\
0 & 0
\end{array}
\right) \quad E_{-}=\left(
\begin{array}{cc}
0 & 0 \\
1 & 0
\end{array}
\right) .
\end{equation}

So that we can calculate $\nu $ as

\begin{equation}
\nu =e^{\frac{1}{2}\phi H}e^{\chi E_{+}}=\left(
\begin{array}{cc}
e^{\frac{1}{2}\phi } & \chi e^{\frac{1}{2}\phi } \\
0 & e^{-\frac{1}{2}\phi }
\end{array}
\right) .
\end{equation}

Also since for $sl(2,\mathbb{R)}$ the generalized transpose is
simply the matrix transpose

\begin{equation}
\mathcal{M=}\nu ^{\intercal }\nu =\left(
\begin{array}{cc}
e^{\phi } & \chi e^{\phi } \\
\chi e^{\phi } & e^{-\phi }+\chi ^{2}e^{\phi }
\end{array}
\right) .
\end{equation}

Now from (2.4) we have the Lagrangian

\begin{equation}
\mathcal{L=-}\frac{1}{2}\ast d\phi \wedge d\phi -\frac{1}{2}e%
^{2\phi }\ast d\chi \wedge d\chi .
\end{equation}

We only have one positive root so $K_{\alpha \beta }^{\gamma }=0$, $%
\omega =0$ and from (2.12) $\mathbf{\Omega =}1.$If we use these results in
(2.15) we have

\begin{equation}
\begin{aligned}
\mathcal{G}_{0}&=d\nu \nu ^{-1}\\
\\
&=\frac{1}{2}d\phi H+e^{\phi }d%
\chi E_{+}.
\end{aligned}
\end{equation}

This could directly be calculated from (5.2).Since $\alpha =2,F=d\chi $
and $N_{1,1}=0$ from (2.19) and (2.20) we have the second-order equations of
motion

\begin{equation}
\begin{aligned}
d(\ast d\phi )&=e^{2\phi }d\chi \wedge \ast d\chi ,\\
\\
d(e^{2\phi }\ast d\chi )&=0.
\end{aligned}
\end{equation}

By using (3.7) the commutation relations for the dual generators can be found as

\begin{gather}
[H,\widetilde{H}]=[E_{+},\widetilde{H}]=0,\notag\\
\notag\\
[H,\widetilde{E}_{+}]=-2\widetilde{E}_{+}\quad, \quad [E_{+},%
\widetilde{E}_{+}]=\frac{1}{2}\widetilde{H}.
\end{gather}

We can now calculate the first-order equations from (4.7).We should first
observe that if we use the commutation relations (5.7) in (3.8) and (3.9) or
directly in (3.3) we find that

\begin{gather}
\widetilde{f}_{11}^{1}=\widetilde{f}_{11}^{2}=\widetilde{f}_{12}^{2}=0\quad,\quad
\widetilde{f}_{12}^{1}=\frac{1}{2},\notag\\
\notag\\
\widetilde{g}_{11}^{1}=\widetilde{g}_{11}^{2}=\widetilde{g}_{12}^{1}=0\quad,\quad
\widetilde{g}_{12}^{2}=-2.
\end{gather}

In order to calculate the first-order equations we will first calculate the exponentials in (4.7).Since $\mathbf{\Gamma }%
_{n}^{k}=\frac{1}{2}\phi ^{i}\,\widetilde{g}_{in}^{k}$ and $\mathbf{\Lambda }%
_{n}^{k}=\chi ^{m}\,\widetilde{f}_{mn}^{k}$ we have the $\mathbf{\Gamma }$ and
$\mathbf{\Lambda }$ matrices as

\begin{equation}
\mathbf{\Gamma =}\left(
\begin{array}{cc}
0 & 0 \\
0 & -\phi
\end{array}
\right)\quad,\quad \mathbf{\Lambda =}\left(
\begin{array}{cc}
0 & \frac{1}{2}\chi  \\
0 & 0
\end{array}
\right) .
\end{equation}

When we exponentiate these matrices and multiply the results we get

\begin{gather}
e^{\mathbf{\Gamma }}=\left(
\begin{array}{cc}
1 & 0 \\
0 & e^{-\phi }
\end{array}
\right)\quad,\quad e^{\mathbf{\Lambda }}=\left(
\begin{array}{cc}
1 & \frac{1}{2}\chi  \\
0 & 1
\end{array}
\right) ,\notag\\
\notag\\
e^{\mathbf{\Gamma }}e^{\mathbf{\Lambda }}=\left(
\begin{array}{cc}
1 & \frac{1}{2}\chi  \\
0 & e^{-\phi }
\end{array}
\right) .
\end{gather}

The vectors in (4.7) are calculated from their definitions given in section
three and section four and by using (5.8)

\begin{equation}
\overset{\rightharpoonup }{\mathbf{\Psi }}=\left(
\begin{array}{c}
\frac{1}{2}d\phi  \\
e^{\phi }d\chi
\end{array}
\right)\quad,\quad \overset{\rightharpoonup }{\mathbf{A}}%
=\left(
\begin{array}{c}
\frac{1}{2}\text{d}\widetilde{\phi } \\
\text{d}\widetilde{\chi }
\end{array}
\right).
\end{equation}

When we substitute these results in the equation (4.7)

\begin{equation}
\left(
\begin{array}{c}
\frac{1}{2}\ast d\phi  \\
e^{\phi }\ast d\chi
\end{array}
\right) =\left(
\begin{array}{cc}
1 & \frac{1}{2}\chi  \\
0 & e^{-\phi }
\end{array}
\right) \left(
\begin{array}{c}
\frac{1}{2}d\widetilde{\phi } \\
d\widetilde{\chi }
\end{array}
\right)
\end{equation}

from which we read the first-order equations of motion for the $SL(2,\mathbb{%
R)}$ /$SO(2)$ coset as

\begin{equation}
\begin{aligned}
\ast d\phi &=d\widetilde{\phi }+\chi d\widetilde{\chi },\\
\\
e^{\phi }\ast d\chi &=e^{-\phi }d\widetilde{\chi }.
\end{aligned}
\end{equation}

These first-order equations are the same equations with the ones in [13]
which are obtained by directly integrating the second-order equations (5.6).

\section{Conclusion}
We have derived the second-order equations of motion for a general scalar coset
sigma model.By following the framework of [13] the general form of the structure constants and the first-order equations are derived for the doubled field strength
formalism.We have shown that these general formulas lead to the correct
results for the case of $SL(2,\mathbb{R)}$ /$SO(2)$.

We have started from any given coset $G/K$ and derived the dualised formulation
as well as the first-order  equations in terms of the given knowledge of the
original global symmetry group of the scalar Lagrangian.The results achieved
in this work are applicable to the scalar sectors of the
maximal supergravities but as they are derived for a general symmetric space
sigma model with the abstract notion of the structure constants
they would dualise and construct the first-order formulation of any other
theory which assumes the same form.Especially the non-linear
realisations of the supergravity theories in various dimensions in which
there is matter coupling would include the scalar formalism introduced here.
The dualisation of the scalar sector is a primary task and a good starting point
for the non-linear realisation of any supergravity theory since it contains the
knowledge of the global symmetry group of the Bosonic sector of the theory.This work
in the most general sense relates the structure of the original scalar coset and the symmetry
group of the doubled field strength from which the first-order equations are derived.

In spite of the fact that we have given the formulation of the coset
sigma models for the maximally non-compact global symmetry group
a similar formulation can be performed for the general case of non-compact groups but in that
case the role of the Borel subgroup must be replaced
by the solvable subgroup.The formulation of the non-linear sigma models and
in general the principle sigma models would follow the same pattern presented here.

\section{Acknowledgements}
This work has been supported by TUBITAK (The Scientific and
Technical Research Council of Turkey).I would like to thank Prof Peter West
and Prof Tekin Dereli for discussions and useful remarks.

\end{document}